\documentstyle[12pt]{article}
\newfont{\capttionfont}{cmr9}
\makeatletter
\long\def\@makecaption#1#2{%
  \vskip\abovecaptionskip
  \sbox\@tempboxa{\capttionfont #1: #2 }%
  \ifdim \wd\@tempboxa > \hsize
   \begin{center} {\capttionfont #1: }\parbox[t]{0.6
\linewidth}{\capttionfont #2}
   \end{center}
  \else
    \hbox to\hsize{\hfil\box\@tempboxa\hfil}%
  \fi
  \vskip\belowcaptionskip}
\makeatother
\begin{document}

\title{Semi-Markov Random Walks and  Universality in Ising-like Chains.}
\author{M. Droz \\
D\'epartement de Physique Th\'eorique \\
Universit\'e de Gen\`eve
CH-1211 Gen\`eve 4 \\ and \\ M.-O. Hongler\\ D\'epartement de
Microtechnique\\ Institut de Production Microtechnique
\thanks{Dirig\'e par le Professeur Jacques Jacot} \\ Ecole Polytechnique
F\'ed\'erale de Lausanne \\ CH-1015 Lausanne}
\maketitle	
\abstract{We exhibit a one to one correspondence between some universal
probabilistic properties of the ordering coordinate of
one-dimensional Ising-like models  and a class of  continuous time
random walks. This correspondence provides an new qualitative picture of
the properties of the ordering coordinate of the Ising model.

%%%%%%%%%%%%%%%%%%%%%%%%%%%%%%%%%%%%%%%%%%%%%%%%%%%%%%%%%%%%%%%%%%%%%

%%%%%%%%%%%%%%%%%%%

\section{Introduction}

Many systems in Nature exhibit a phase transition when a control
parameter (the reduced temperature for example) $\epsilon$ is varied.
Often, the phase transition is related to a spontaneous symmetry
breaking in the system: one phase is less symmetric than the other, and
the so-called order parameter $\phi$ describes this symmetry breaking.
The parameter $\phi$ could be a scalar, a vector or a tensor and its
number of  components $n$  is an important parameter.

The amplitude of the order parameter is zero in the most symmetric phase
and nonzero in the less symmetric one.
If the order parameter varies continuously at the transition, one speaks
of a second order phase transition.
It turns out that, near a second order transition point, every physical
quantity $\cal A$ (like the specific heat, the susceptibility, ...) can be
written as the sum of a regular
part and a singular one in $\epsilon$. The singular part (${\cal A}_s$)
behaves  as power law ${\cal A}_s \sim {\vert \epsilon  \vert}^{-\alpha}$
providing that the control parameter $\epsilon$ has been chosen as
to vanish at the critical point. The parameter $\alpha $ is called the
``critical
exponent'' associated to  $\cal A$.

A remarquable feature of second order phase transitions is the
so-called ``universality''. Scaling theory [1] and renormalization group
arguments [2] show that, the critical exponents are
only function of $n$ and $d$, the dimensionality of the system. Thus
systems as different as for example an Ising model (describing the
paramagnetic ferromagnetic transition in one component spin system) and
a model of a fluid (describing  the liquid gas transition)  belong to
the same universality class.

A less investigated problem  concerns the universal aspect of ordering
coordinate underlying critical point singularities.
A. Bruce [3] has investigated this problem in the case of an Ising-like
problem, i.e. a problem for which the scalar order parameter is the
ensemble average of a one-component local spin variable $\sigma(x)$. The
configurational energy consists of local double well potentials and the
short-range interaction among the spins.
Each stochastic variable $\sigma(x)$ is characterized  by its probability
density distribution $P(\sigma)$. This function has a whole spectrum
of possible forms, but two extreme cases can be singularized. At low
temperature, the thermal energy is much smaller
than the depth of the potential wells. Thus the ordering coordinate are
localized in the vicinity of the two well minima
and $P(\sigma)$ has a strongly two-peaked form. One speaks of the
``ordered'' limit. Alternatively, at high temperature,
the potential wells are so shallow that the local coordinate is nearly
distributed about the symmetric state $\sigma=0$. One speaks of the
``disordered'' limit.

%%%%%%%%%%%%%%%%%%%%%%%%%%%%%%%%%%%%%%%%%%%%%%%%%%%%%%%%%%%%%%%%%%%%
Parallel to the  spin system, the continuous time random walks, (also
called randomized random walks in W. Feller, [4]), do play an important
role in the modeling of transport phenomena in disordered media, (see
[5], for a recent  bibliography on this topic). Firm mathematical basis
to  describe this class of processes, can be found in [6] and [7], the
later being  more directly related to one aspect of the models to be
described here. The simplest situation consists of a symmetric random
walks on the infinite one-dimensional, ($1-D$),  lattice with lattice
spacing $a$ and with the random time between consecutive the jumps
being exponentially distributed. In this case, the position of the
walker is described by a markovian Master Equation.  When the time
between the jumps is drawn from   non-exponential distributions,
non-markovian properties arise and to describe these situations E.
Montroll and G. Weiss proposed  a generalization of the Master equation
[8]. In the present paper, we shall consider the non-markovian random
walks obtained  when  the time intervals between jumps are drawn from
generalized Erlang distributions. These distributions defined on $[0,
\infty]$ belong to the general class of phase-type distributions which
describe  the time until absorption of a finite state continuous time
Markov chain [9]. For this class of random processes, we shall observe
that, when large time are considered, the probablistic properties of
the position of the walker  exhibit a universal behavior which is in one
to one correspondence with the probabilistic properties of the ordering
coordinates of a one-dimensional Ising spin systems.

Our paper is organized as follows in section 2, a brief account of the
universal properties of Ising-like
systems are given. In particular, the probability distribution of a
collective
coordinates in an Ising chain is recalled. In section 3,  we introduce
the relevant class
of continuous time random walks  to be used and calculate the probability
density describing the position of the walker. In the last section, we
exhibit the one to one correspondence which can be drawn between the two
classes of models.
%%%%%%%%%%%%%%%

\section{Universal properties of the probability density distribution of
the collective  coordinates in  Ising-like models.}

Let us return to the Ising-like system  described in the introduction.
To be able to detect the universal features of the ordering coordinates,
one
has then to coarse-grained the problem and considers ``bloc-spin''
variables $\sigma_L(x)$ representing the effective spin of a block of
linear size $L$. The variable $\sigma_L$ is also a random variable
characterized by
its  probability density distribution $P(\sigma_L,L)$.
Renormalization group arguments suggest that $P_L(\sigma_L,L)$ tends to an
universal  limiting form $P^*$ when both $L$ and the correlation length
$\xi$ are much larger than $a$, the lattice spacing.
It turns out that in one dimension the probability density function $P^*$
can be computed exactly using a matrix transfert operator technique [3].
Furthermore, the critical temperature of one-dimensional Ising-like
systems
with short range interactions is zero. The low temperature properties of
the block coordinate probability distribution function, for large block
size, is governed by the statistical mechanics of the cluster walls [10].
Thus to compute  $P(\sigma_L,L)$ for our one-dimensional Ising like
models, it
suffices [3] to consider a one dimensional fixed-length spin Ising model,
$\sigma(x_i)=\pm \sigma_0$,
with nearest neighbors interactions and the following effective
Hamiltonian (in units of $k_BT$, where $k_B$ is the Boltzmann constant and
$T$ the temperature): \begin{equation}
{\cal E}=-\frac{K}{\sigma_0^2} \sum_{i=1}^{N} \sigma(x_i) \sigma(x_{i+1})
\end{equation}
$\sigma_0$ is chosen such as $P(\sigma_L,L)$ has an unit variance.

\noindent
It is known [11] that (in the thermodynamic limit $N \to \infty)$ the 
two-point correlation function  of this  model is:
\begin{equation}
C_2(r)=<\sigma(x)\sigma(x+r)>=\sigma_0^2~ [{\rm tanh}(K)]^r=\sigma_0^2
\exp(-r/\xi)
\label{TANK}
\end{equation}
where
\begin{equation}
\xi=\frac{1}{2}\exp(2K)[1+{\cal O}(\exp(-4K))]
\end{equation}
is the correlation length.

\noindent
The block coordinate takes the simple form
\begin{equation}
\sigma_L=\frac{1}{L\sigma_0}\sum_{i=1}^{L} \sigma(x_i).
\end{equation}

\noindent
To compute $P(\sigma_L,L)$ it is suitable to introduce its Fourier transform ${\tilde P}(H,L)$:
\begin{equation}
P(\sigma_L,L) = \frac{1}{2\pi}\int_{-\infty}^{\infty} {\tilde P}(H,L)
\exp(-iH\sigma_L)dH,
\end{equation}
${\tilde P}(H,L)$ is simply the characteristic function from which the
n-point cumulants can be derived. Moreover, ${\tilde P}(H,L)$ can also be
written as [3]:
\begin{equation}
{\tilde P}(H,L) =\frac{Z(iH)}{Z(0)},
\end{equation}
where $Z(h)$ is the canonical partition function defined as
\begin{equation}
Z(h)= Tr_{\{\sigma(x_i) \} } \exp [-{\cal E}
+\frac{h}{L\sigma_0}\sum_{i=1}^{N}
\sigma(x_i) ].
\end{equation}
The traces $Z(h)$ and $Z(0)$ can be readily computed by the usual transfert
matrix
method [11] for $h$ real.
The analytic continuation to the  imaginary $h$ axe is unique.

\noindent
In the large $L$ and $\xi$ regime, one finds [3] that  ${\tilde P}(H,L)$
is indeed an universal function, denoted ${\tilde P}^*(H,z)$, depending
only upon the reduced variable $z=L/\xi$:
\begin{equation}
	{\tilde P}^*(H,z)= \exp(- {z\over 2})\left\{
	\cosh \left(\sqrt{{z^{2} \over 4}-H^{2}}\right)+ {1 \over
	\sqrt{1-{4H^{2} \over z^{2}}}}
	 \sinh \left(\sqrt{{z^{2} \over 4}-H^{2}}\right)
	\right\}.
	\label{bru1}
\end{equation}

\noindent
By inverse Fourier transform one finds:
\vfill\eject
$$
P^*(\sigma_L, z)={1 \over 2}\exp(-{z\over 2}) \Big\{ \left(
\delta(1-\sigma_L)
+\delta(1+\sigma_L) \right) \Big\}+
$$
\begin{equation}
{z \over 4}\,\exp(-{z\over 2})
\Big[ {\rm I}_0 \left(\frac{z}{ 2}{(1-\sigma_L^2)}^{1/2} \right)+
\frac{ {\rm I}_1 \left(\frac{z}{ 2}{(1-\sigma_L^2)}^{1/2}
\right)}{{(1-\sigma_L^2)}^{1/2}}
\Big] \Theta(1-\sigma_L^2). \label{bru}
\end{equation}
Two particular limits deserve a special attention:

\begin{itemize}
	\item  i). {\sl The critical limit:} $z \to 0$.

In this case one finds:
\begin{equation}
P^*(\sigma_L,z) \approx \frac{1}{2}\Big[
\delta(1-\sigma_L)+\delta(1+\sigma_L) \Big].
\label{LIM1}
\end{equation}
Hence the  probability distribution function has a limiting ``ordered''
form.

\item  ii). {\sl The non-critical limit:} $z \to \infty$.

In this cas, one finds
\begin{equation}
P^*(\sigma_L,z) \approx \sqrt{\frac{z}{4\pi}} \exp
\Big(\frac{-z\sigma_L^2}{4}    \Big)
\label{LIM2}
\end{equation}
which corresponds to the ``disordered'' limit.
\end{itemize}
%%%%%%%%%%%%%%%%%%%%%%%%%%%%%%%%%%%%%%%%%%%%%%%%%%%%%%%%%%%%%%%%%%%%%%%

\section{Semi-Markov Birth and Death processes}
Now we consider a symmetric birth and death process on the infinite $1-D$
lattice:

\noindent
${\rm Z}_{a}= \left\{...,-a,0,+a, 2a,...\right\}$, $a$ being the
lattice spacing. The jumps of the random walker occur at random times
$\xi_{0},\, \xi_{1}, \,\xi_{2}$ and the interval of time between
successive jumps is denoted $\tau_{k}= \xi_{k+1}-\xi_{k}$ , $k=0,\,
1, \,2, ...$ . We assume the $\tau_{k}$ to be independent and identically
distributed random variables. Let $\psi(\vec{\lambda}, \vec{\mu}, t)$ be
the probability density from which the
$\left\{\tau_{k}\right\}$'s  are drawn. The constant vectors
$\vec{\lambda}=\left\{\lambda_{0},\lambda_{1},....\lambda_{N} \right\}$ and
$\vec{\mu}=\left\{\mu_{0}, \mu_{1},...\mu_{M}\right\}$ are parameters
characterizing the distributions of the time between consecutive jumps.
Accordingly, we shall write:

\begin{equation}
	{\rm Prob}\left\{t \leq \tau \leq t+dt\right\}=\psi(\vec{\lambda},
	\vec{\mu}, t) dt.
	\label{DEF1}
\end{equation}
From now on, we shall assume that $\psi(\vec{\lambda}, \vec{\mu}, t)$ is
called
a PH-Type density, which can be defined by its Laplace transform:

\begin{equation}
	\int_{0}^{\infty}\psi(\vec{\lambda}, \vec{\mu}, t) e^{-ut} dt =
	\widetilde{\psi}(\vec{\lambda}, \vec{\mu}, u)={\Pi_{\vec
	{\mu}}\over\Pi_{\vec {\lambda}}}
	\label{PHASE}
\end{equation}
with the definitions:

\begin{equation}
		\Pi_{\vec {\lambda}}= \left\{\begin{array}{lll}
		\prod_{k=1}^{N}(1 + {u \over \lambda_{k}}),      
& \mbox{for $N  \geq 1
$,}   \\
		\\
		1,  &\mbox{for $ N=0 $.}
		\end{array} \right.
	\label{DEF2}
\end{equation}
and
\begin{equation}
		\Pi_{\vec {\mu}}= \left\{\begin{array}{lll}
		\prod_{k=1}^{M}(1 + {u \over \mu_{k}}),      & \mbox{for $M  \geq 1
$,}   \\
		\\
		1,  &\mbox{for $ M=0 $.}
		\end{array} \right.
	\label{DEF3}
\end{equation}

\noindent
We impose that $N>M$. In view of Eqs.(\ref{PHASE}), (\ref{DEF2})
and (\ref{DEF3}), we immediately obtain :
\begin{equation}
	\langle \tau \rangle=
	- {\partial \over \partial u}\widetilde{\psi}(\vec{\lambda}, 
\vec{\mu},
u)\mid_{u=0}
	=\sum_{k=0}^{N}{1 \over \lambda_{k}}-\sum_{k=0}^{M}{1 \over \mu_{k}}>0,
	\label{AVERAGE}
\end{equation}
and the variance of the time between consecutive jumps reads as:
$$
	\sigma_{\tau}^{2}=\langle \tau^{2} \rangle - \langle \tau \rangle^{2}=
		 {\partial^{2} \over \partial u^{2}}\widetilde{\psi}(\vec{\lambda},
		 \vec{\mu}, u)\mid_{u=0}-
	\left[- {\partial \over \partial
	u}\widetilde{\psi}(\vec{\lambda}, \vec{\mu}, u)\mid_{u=0}\right]^{2}=
	$$
	\begin{equation}
	 \left[\sum_{k=1}^{N}{1 \over \lambda_{k}}\right]^{2}-\sum_{k \neq m}^{N}
{1 \over \lambda_{k} \lambda_{m}}-
\left[\sum_{k=1}^{M}{1 \over \mu_{k}}\right]^{2}+\sum_{k \neq m}^{M} {1
\over \mu_{k} \mu_{m}}.
	\label{VARIATION}
	\end{equation}
	
\noindent
Let us now return to our random walk and denote by $P(sa,t)$ the
probability to find the walker at the position $sa$ at time $t$. We shall
assume:
\begin{equation}
	P(sa,0)= \delta_{sa,0}\,\,\,\,\,{\rm and} \,\,\,\,\,\xi_{0}=0.
	\label{INIT}
\end{equation}
where $\delta_{sa,0}$ is the Kronecker symbol. Using the formalism
introduced in [5] and [6], it is shown in [12] that
$P(sa,t)$ obeys to the high order differential difference equation:
\begin{equation}
\prod_{k=0}^{N}(1+ {1 \over \lambda_{k}}{\partial \over \partial t})
P(sa,t) = \prod_{k=0}^{M}(1+ {1 \over \mu_{k}}{\partial \over \partial
t})\left[P\left((s+1)a, t\right)+ P\left((s-1)a, t\right)\right].
	\label{CHAPMANN}
\end{equation}
To solve Eq.(\ref{CHAPMANN}), we would need to specify the appropriate
initial
conditions, ${\partial^{m} \over \partial
t^{m}}P(sa,t) \mid_{t=0}$ for $m=1,2,...,N-1$.  As we shall not derive, in
this paper,  solutions of Eq.(\ref{CHAPMANN}), we refrain to
give here further details on the characterization of these initial
conditions.

\noindent
Let us now give two simple illustrations:

\begin{itemize}
	\item  \underline{Example 1}: With $N=1$ and $M=0$, 
Eq.(\ref{CHAPMANN}) takes the form:
\begin{equation}
	\left(1+ {1 \over \lambda_{0}}{\partial \over \partial t}
\right)P(sa,t)=
	{1 \over 2}\left[P\left((s+1)a, t\right)+
 P\left((s-1)a, t\right)\right]
	\label{EX}
\end{equation}
which is the Master equation for the continous time Markov random walk.
	
	\item
\underline{Example 2}: With $N=2$, $\lambda_{0} = \lambda_{1}$ and $M=0$,
Eq.(\ref{CHAPMANN}) takes the
form:
$$
	\left(
	{\partial^{2} \over \partial t^{2}}P(sa,t)+ 2 \lambda_{0}
{\partial \over \partial t}P(sa,t)
	\right)P(sa,t)=
	$$
	\begin{equation}
	 \lambda^{2}_{0}\left[- P(sa, t ) +{1 \over 2}P\left((s+1)a, t\right)+
	 {1 \over 2}P\left((s-1)a,t\right)\right]
	\label{EX2}
\end{equation}
\noindent
with the initial condition  being for this case: ${\partial \over \partial
t}P(sa,t) \mid _{t=0}$.
\end{itemize}

\noindent
For small lattice spacing $a$, we expand the right hand side of
Eq.(\ref{CHAPMANN}) up to second order in $a$:
\begin{equation}
	\prod_{k=1}^{N}\left(1 + {1 \over \lambda_{k}}{\partial \over \partial
t}\right)P(x, t)=
	\prod_{k=1}^{M}\left(1 + {1 \over \mu_{k}}{\partial \over \partial
	t}\right)\left[P(x, t) + {a^{2} \over 2}{\partial^{2} \over \partial
	x^{2}}P(x,t)\right]
	\label{LAPLACE}
\end{equation}
%%%%%%%%%%%%%%%%%%%%%%%%%%
where we have written $sa=x \in {\cal R}$. As usual, we now rescale  the
time variable as:
\begin{equation}
	T = a^{2} t.
	\label{scale}
\end{equation}
Introducing Eq.(\ref{scale}) into Eq.(\ref{LAPLACE}), we obtain:
$$\sum_{k>2}^{N} {\cal C}_{k}a^{2k}{\partial^{k} \over \partial T^{k}} P(x,
T) +
	a^{4} {\cal C}_{2} {\partial ^{2} \over \partial T^{2}} P(x, T) +
 a^{2} {\cal C}_{1} {\partial \over
	\partial T}P(x,T)=
	$$
	\begin{equation}
		{a^{2} \over 2} {\partial ^{2} \over \partial x^{2}}P(x, T) + \sum_{k
		\geq 1}^{M} a^{2k+2} {\cal B}_{k}{\partial^{k} \over \partial
T^{k}}{\partial^{2}
		\over \partial x^{2}}P(x,T),
	\label{FULLTEL}
\end{equation}
where ${\cal C}_{k}$ and  ${\cal B}_{k}$  are constant coefficients
$\forall k$. Using Eq.(\ref{AVERAGE}),  we have in particular:
\begin{equation}
	{\cal C}_{1} = \langle \tau \rangle.
	\label{C1}
\end{equation}

\noindent
When $M=0$, the PH-Type distribution  Eq.(\ref{PHASE}) reduces to the
${\rm E}_{n}$-type distribution, (generalized Erlang distributions of
order $n$). From now on, we shall focus our attention to this
class of ${\rm E}_{n}$-type models. Retaining the terms up to order
$a^{2}$ in
Eq.(\ref{FULLTEL}), we have therefore  when $M=0$:
\begin{equation}
	a^{2} {\cal C}_{2} {\partial^{2} \over \partial T^{2}} P^*(x, T) +
\langle \tau
	\rangle {\partial  \over \partial T}P^*(x, T)=
 {1 \over 2}{\partial ^{2}
	\over \partial x^{2}} P^*(x, T).
	\label{TEL}
\end{equation}
\noindent
where the superscript $*$ for $P^*(x, T)$ indicates that we restrict
ourselves to the  ${\rm E}_{n}$-type model. In view of Eqs.
(\ref{FULLTEL}),
(\ref{AVERAGE}) and (\ref{VARIATION}), we can write:
\begin{equation}
	{\cal C}_{2}= {1 \over 2} \left(\langle \tau \rangle ^{2}
	-\sigma_{\tau}^{2}\right)={1 \over 2} \langle \tau \rangle ^{2}
\left(1- CV_{\tau}^{2}\right)
	\label{C2}
\end{equation}
where ${\rm CV}_{\tau}^{2}$ stands for the coefficient of variation which
for
${\rm E}_{n}$-type distribution is always smaller than unity.

\noindent
Eq.(\ref{TEL}) is the celebrated Telegrapher's equation which, in the
asymptotic regime,  appears as a  "universal" description for all random
walks  with time
between the jumps drawn from  ${\rm E}_{n}$-type distributions.

\noindent
Introducing the Fourier transform:
\begin{equation}
	\widetilde{P}^*(k,T)= \int_{0}^{\infty} P^*(x, T) \, e^{-ikx} \, dx,
	\label{FOURIER}
\end{equation}
the Telegrapher's Eq.(\ref{TEL}) takes the form:
\begin{equation}
	a^{2} {\cal C}_{2} {\partial^{2} \over \partial T^{2}} P^*(k, T) +
\langle \tau
	\rangle {\partial  \over \partial T}P^*(k, T)=
 {k^{2} \over 2} P^*(k, T)
	\label{FTEL}
\end{equation}
and the solution of Eq.(\ref{FTEL}) with initial conditions:
\begin{equation}
	\widetilde{P}^*(k, T=0) = 0 \,\,\,\,\,\,{\rm and}\,\,\,\,\,\, 
{\partial \over \partial
	T}\widetilde{P}^*(k, T) \mid _{T=0}=0
	\label{IC}
\end{equation}
%%%%%%%%%%%%%%%%%%%%%%%%
reads immediately as in Eq.(\ref{bru1}), provided we  reinterpret
the
parameters as:

\begin{equation}
	z= {2 T \over q \langle \tau \rangle a^{2}} \,\,\,\,\,{\rm
	and}\,\,\,\,\, H^{2}= {q a^{2}\over 4} z^{2} k^{2},
	\label{NN}
\end{equation}
with
\begin{equation}
	q=(1-{\rm CV}_{\tau}^{2}) \,\,\in \,\,[0,1].
	\label{q}
\end{equation}
By direct inversion, the density Eq.(\ref{bru}) follows provided we
introduce the identification:
\begin{equation}
	\sigma_{L}^{2}={4x^{2} \over  z^{2}a^{2}q}.
	\label{ident}
\end{equation}

\noindent
Although this is not necessary, we can chose $T=1$ which corresponds to the
definite scale factor  chosen in [3], (see Eq. (2.11) in A. Bruce, [3]).

\noindent
Hence, the critical and non critical regimes described by Eq.(\ref{LIM1})
and (\ref{LIM2}) follows directly.

%%%%%%%%%%%%%%%%%%%%%%%%%%%%%%%%%%%%%%%%%%%%%%%%%%%%%%%%%%

\section{Conclusion}

The Telegrapher's equation Eq.(\ref{TEL}) and its solution
Eq.(\ref{bru}) describe the asymptotic  behavior  of the probability
density for  both the coarse-grained variables of an
Ising-like  chain and the position of continuous time random walker.

The behavior of the block coordinate of the Ising system is similar to the
behavior of a random walker with time between the jumps drawn from an
$E_n$ type distribution. This can be understood qualitatively as follows:
consider the coarse-grained variable of the Ising model including a
given, (large), number of spins, say $N >> 1$. Add one spin to this sample
of $N$ spins
and consider the new probability density distribution of the block spin
coordinate. Two main factors influence the result:

\begin{itemize}
	\item  1) the correlation between nearest-neighbor spins.
	
	\item  2) the fact that in the large sample $N$ limit, only the spins
with values close to $\pm \sigma_{0}$ will contribute to the
coarse-grained variables.
\end{itemize}

\noindent
Hence, to an increase of the sample used to construct the coarse-grained
spin variable in the Ising model corresponds an increase of the time in
the continuous time random walk with a lattice sapcing being related to
$\sigma_{0}$.
Furthermore,  the "aging" effect implied by the use of non-exponential
distributions for the time intervals between the jumps directly describes
the correlation
between the spin variables in the Ising model. Quantitatively,
the  roles played by the parameters of the two models  are as follows:
to the parameter ${\rm tanh}(K) \in [0,1], \,\,\,\, K \geq 0$ defined in
Eq.(\ref{TANK}), we can
directly  associate the parameter
$q \in [0,1]$ defined in Eq.(\ref{q}). This correspondence is physically
consistent as  ${\rm tanh}(K)$  measures the  correlation between
nearest-neighbor spin
variables given  by Eq.(\ref{TANK}) while $q$ relates the memory effects of
the semi-markovian random walk. Two limiting regimes deserve attention:

\begin{itemize}

	\item  a) \underline{Weak correlation regime.} Here $q \rightarrow 0 \,\,
\Leftrightarrow\,\, CV_{\tau}^{2}=1$ and hence ${\rm tanh}(K) \Leftrightarrow
0 \Rightarrow K \,\, \rightarrow 0 $. We then  reach here the markovian
regime of the random walk. Accordingly, the $z \rightarrow
\infty$ limit arises in Eq.(\ref{NN}) which implies an absence of
correlations
between the spins.
	
	\item  b) \underline{Strong correlation regime.} To the limit $K
\rightarrow \infty
\,\,\Rightarrow \,\,{\rm tanh}(K) \rightarrow 1$ corresponds  the  $q=1
\Rightarrow  CV^{2}_{\tau} =0$ regime for the continuous time random walk.
As
$CV^{2}_{\tau}=0$, the time intervals between the jumps become equally
spaced and this therefore  corresponds  to a discrete, (deterministic),
time random walk.
Accordingly, we have $z \rightarrow 0$ and therefore the double-peak shape
of the
probability density  exhibited in Eq.(\ref{LIM1}) arises consistently.
\end{itemize}

%%%%%%%%%%%%%%%%%%%%%%%%%%%%%%%%%%%%%%%%%%%%%%%%%%

\section*{Acknowledgments}

MD is partially supported by the Swiss National Science Foundation.

\noindent
\section*  {References}

\vspace {0.5 cm}
\noindent
[1] Stanley H.E. 1971 {\it Introduction to Critical Phenomena}, (Oxford:
Clarendon)

\vspace {0.5 cm}
\noindent
[2] Ma S.K. 1976 {\it Modern Theory of Critical Phenomena}, (Reading,
Mass: Benjamin)

\vspace {0.5 cm}
\noindent
[3] Bruce A. D. "Probability density Functions for Collective coordinates
in Ising-Like Systems" 1981 J. Phys. C: Soliod State Phys. {\bf 14}
3667-3688

\vspace {0.5 cm}
\noindent
[4] Feller W. {\it An introduction to Probability theory and its
applications} 1971, Vol.  II,  (J. Wiley, New-York).

\vspace {0.5 cm}
\noindent
[5] Hughes B. D. {\it Random Walk and Random Environments} 1995, Vol. II,
(Clarendon Press, Oxford)

\vspace {0.5 cm}
\noindent
[6] Cinlar E. "Markov Renewal Theory: A Survey" 1975 Management Science
{\bf 21}, 727-752.

\vspace {0.5 cm}
\noindent
[7] G. Latouche. "A Phase Type Semi-Markov Point process".  1982 SIAM J.
Alg. Disc. Meth. {\bf 3}, 77-90.

\vspace {0.5 cm}
\noindent
[8] Montroll E. W. and   Weiss G. H. "Random Walks on Lattice II" (1965) J
of Math. Phys. {\bf 6},
167-181.

\vspace {0.5 cm}
\noindent
[9] Neuts M. 1981 {\it Matrix Geometric Solutions in Stochastic Models}
(John  Hopkins Press, Baltimore)

\vspace {0.5 cm}
\noindent
[10] Krumhansl J.A. and Schreiffer J.R. "Dynamics and statistical
mechanics of a one-dimensional model Hamiltonian for structural phase
transitions"  1975 Phys. Rev. {\bf B11} 3535-3545.

\vspace {0.5 cm}
\noindent
[11] Thompson C.J. 1972 {\it Mathematical Statistical Mechanics},
(Princeton Univ. Press).

\vspace {0.5 cm}
\noindent
[12] Hongler M.-O. and Salama Y. "Semi-Markov Processes with Phase-Type
waiting times"  1996 Zeit. fr Angew. Math. und Mech. (ZAMM) {\bf 76}
Suppl. 461-462

\end{document}